\documentstyle[12pt]{article}
\input{epsf}
\newcommand{\be}{\begin{eqnarray}}
\newcommand{\ee}{\end{eqnarray}}
\newcommand{\ba}{\begin{array}}
\newcommand{\ea}{\end{array}}

\begin{document}
\renewcommand{\thefootnote}{\fnsymbol{footnote}}
%
%

\rightline{RUB-TP2-12/02}
\vspace{0.5cm}
\begin{center}
{\Large On ``dual" parametrizations of generalized parton distributions}\\
\vspace{0.35cm}
 M.V. Polyakov$^{a,b}$, A.G.~Shuvaev$^{a}$\\

\vspace{0.35cm}
$^a$Petersburg Nuclear Physics
Institute, Gatchina, St.\ Petersburg 188350, Russia\\
and\\
$^b$Institut f\"ur Theoretische Physik II,
Ruhr--Universit\"at Bochum, D--44780 Bochum, Germany

%
%
\end{center}

\begin{abstract}
We propose a parametrization for the generalized parton
distributions (GPDs) which is based on representation of parton
distributions as an infinite series of $t$-channel exchanges. The
entire generalized parton distribution is given as an infinite sum
over contributions of generalized light-cone distribution
amplitudes in the $t$-channel. We also discuss the relations of
the lowest Mellin moments of GPDs to basic mechanical
characteristics of the nucleon as a compound system.
\end{abstract}
\vspace{0.1cm}

\section*{\normalsize \bf  Introduction}
After the first experimental data on deeply virtual Compton
scattering (DVCS) \cite{ZEUS,H1,Amarian,JLAB} and hard exclusive
meson production \cite{HERAmeson,HERMESmeson} have been presented,
it is the right time to address the problem of the extraction of
the twist-2 generalized parton distributions (GPDs) from the
observables. One of the possibilities to do this is to introduce a
parametrization of GPDs and then to extract the corresponding
parameters from the data. The parametrization should respect all
general constrains for GPDs (for reviews see
\cite{Jirev,Radrev,GPV,Belitsky:2001ns}) and it
should have clear physical interpretation.

To our best knowledge in all calculations of observables for hard
exclusive processes the parametrization for GPDs based on double
distributions (DDs) \cite{DM,Rad96,Rad} has been used. In this
parametrization
(supplemented by the contribution of the D-term \cite{PW99})
an important property of polynomiality of the Mellin moments \cite{Ji} of GPDs is
satisfied automatically. The double distributions are modeled
usually by the ansatz of the form \cite{Rad} (for details see next section):

\be
\label{Radparam}
F(\beta,\alpha)=h(\alpha,\beta)\
q(\beta)\, .
\ee
Here $q(\beta)$ is usual quark distribution and the profile
function $h(\alpha,\beta)$ popularly is chosen usually in the form
\cite{Rad}:

\be
h(\alpha,\beta)=\frac{\Gamma\left(2 b+2\right)}{2^{2b+1}\Gamma^2\left(b+1\right)}
\frac{\left[(1-|\beta|)^2-\alpha^2\right]^b}{(1-|\beta|)^{2b+1}}\,.
\ee
Here
the parameter $b$
characterizes the strength of the $\xi$ dependence of the
resulting GPDs. The advantage of the ansatz (\ref{Radparam}) is
its simplicity, however
this ansatz is too
restrictive to have enough flexibility in modeling of GPDs. Also this
form of DD does not commute with the QCD evolution, $i.e.$ if we
assume this ansatz at one normalization point than at higher normalization
points it is generically impossible to put the resulting DD into
the form given by eq.~(\ref{Radparam}) \cite{Mus}.

In the present paper we suggest alternative way to parametrize
GPDs based on the partial wave expansion of the GPDs. Before
the discussion of the new parametrization we present in the next
two sections the discussion of important property of polynomiality of
Mellin moments and  the relations  of the lowest Mellin moments to the mechanical properties of the nucleon.

In this notes we shall restrict ourselves only to the singlet quark GPDs.
The formulae relevant to the
nonsinglet GPDs are collected in Appendix~A. We also do not
consider the gluon GPDs because their consideration is very close
in spirit to that of the quark singlet GPDs.
For GPDs we use notations of X.~Ji (see e.g. \cite{Jirev}).
Generically GPDs are functions of three variable (the fourth is log-dependence on the scale)
$H(x,\xi, t)$, to simplify notations we shall omit the variable $t$ in the case when
a quantity is assumed to be at $t=0$ or when the $t$-dependence is irrelevant for the discussion.

\section*{\normalsize \bf Digression about polynomiality of Mellin moments and D-term}

In this section we do not write explicitly the $t$ dependence of the GPDs
as irrelevant for the present discussion but we always assume it.
One of the non-trivial properties of the generalized parton
distributions is the polynomiality of their Mellin moments.
The polynomiality property means that \cite{Ji}

\be
\label{pc}
\int_{-1}^1 dx\ x^N\ H(x,\xi)&=&h_0^{(N)}+h_2^{(N)}\ \xi^2+\ldots+ h_{N+1}^{(N)}\
\xi^{N+1}\, ,\\
\nonumber
\int_{-1}^1 dx\ x^N\ E(x,\xi)&=&e_0^{(N)}+e_2^{(N)}\ \xi^2+\ldots+ e_{N+1}^{(N)}\
\xi^{N+1}\, .
\ee
Due to the fact that the nucleon has spin $1/2$,
the coefficients in front of the highest power of
$\xi$ for the functions $H$ and $E$ are related to each other \cite{Ji}:
\be
\label{HE}
e_{N+1}^{(N)}=-h_{N+1}^{(N)}\, .
\ee

The polynomiality conditions (\ref{pc}) strongly restrict the
class of functions of two variables
$H(x,\xi)$ and $E(x,\xi)$. For example the conditions (\ref{pc})
imply that GPDs should satisfy the following integral constrains \cite{GPV}:
\be
\label{crit}
\int_{-1}^1 \frac{dx}{x} \biggl[H(x,\xi+x z)-H(x,\xi) \biggr]=
-\int_{-1}^1 \frac{dx}{x} \biggl[E(x,\xi+x z)-E(x,\xi)
\biggr]=z \sum_{n=0}^\infty h_{n+1}^{(n)}\ z^n \, .
\ee
Note that the skewedness parameter $\xi$ enters the lhs of this
equation, whereas the rhs of the equation is $\xi$-independent.
Therefore this $\xi$-independence of the above integrals is a
criterion of whether functions $H(x,\xi)$, $E(x,\xi)$ satisfy the
polynomiality conditions (\ref{pc}). Simultaneously these
integrals are generating functions for the highest coefficients
$h_{N+1}^{(N)}$. Also the condition (\ref{crit}) shows that there
are nontrivial functional relations between functions $H(x,\xi)$
and $E(x,\xi)$, more on this see below.

An elegant possibility to implement the polynomiality conditions
(\ref{pc}) for the GPDs is to use the double distributions
\cite{DM,Rad96,Rad}.
In this case the generalized distributions
are obtained as a one--dimensional section of the two--variable
double distributions $F(\beta,\alpha)$ and $K(\beta,\alpha)$:

\be
\label{dd2}
H(x,\xi)=
\int_{-1}^{1}d\beta\
\int_{-1+|\beta|}^{1-|\beta|} d\alpha\
\delta(x-\beta-\alpha\xi)\  F(\beta,\alpha)\ \, ,
\ee
and an analogous formula for the GPD $E(x,\xi)$:
\be
\label{dd2e}
E(x,\xi)=
\int_{-1}^{1}d\beta\
\int_{-1+|\beta|}^{1-|\beta|} d\alpha\
\delta(x-\beta-\alpha\xi)\  K(\beta,\alpha)\ \, .
\ee
[Recently
in refs.~\cite{BMKS,OlegRadon}
the inversion formula
has been derived]

It is easy to check that the GPDs obtained by reduction from the
double distributions satisfy the polynomiality conditions
(\ref{pc}) but always with $h_{N+1}^{(N)}=e_{N+1}^{(N)}=0$, {\em i.e.}
the highest power of $\xi$ is absent. In other words the
parametrization of GPDs in terms of double distributions is not
complete. It can be completed adding the so-called D-term to
eq.~(\ref{dd2}) \cite{PW99}:

\be
H(x,\xi)&=&
\int_{-1}^{1}d\beta\
\int_{-1+|\beta|}^{1-|\beta|} d\alpha\
\delta(x-\beta-\alpha\xi)\  F(\beta,\alpha)+
\theta\left[1-\frac{x^2}{\xi^2}\right]\ D\left(\frac{x}{\xi}\right)\ \, ,\\
E(x,\xi)&=&
\int_{-1}^{1}d\beta\
\int_{-1+|\beta|}^{1-|\beta|} d\alpha\
\delta(x-\beta-\alpha\xi)\  K(\beta,\alpha)-
\theta\left[1-\frac{x^2}{\xi^2}\right]\ D\left(\frac{x}{\xi}\right)\ \, .
\ee
Here $D(z)$ is odd function having a support $-1\le z\le
1$. In the Mellin moments the D-term generates the highest power
of $\xi$:

\be
h_{N+1}^{(N)}=-e_{N+1}^{(N)}=\int_{-1}^1 dz\ z^N\ D(z)\, .
\ee
Note that for both GPDs $H(x,\xi)$ and $E(x,\xi)$ the D-term is
the same, it contributes to both functions with opposite signs.

The D-term evolves with the change of the renormalization scale
according to ERBL \cite{ER,BL} evolution equation. Hence it is
useful to decompose the D-term in Gegenbauer series (eigenfunctions
of the LO ERBL evolution equation):
\be
\label{dterm}
D(z)&=& (1-z^2) \biggl[ d_1 \ C_1^{3/2}(z)
+ d_3\ C_3^{3/2}(z) + d_5 \ C_5^{3/2}(z) + ...\biggr] \, ,\\
\label{dtermg}
D_g(z)&=& \frac 34 (1-z^2)^2 \biggl[ d_1^G
+ d_3^G\ C_2^{5/2}(z) + d_5^G \ C_4^{5/2}(z) + ...\biggr] \, .
\ee

\section*{\normalsize \bf GPDs and ``mechanical properties" of the nucleon}

The $x$-moments of the GPDs $H$ and $E$ play a special role as
they are related to the form factors of the symmetric energy
momentum tensor. The nucleon matrix element of the traceless part
of the
symmetric energy momentum tensor is characterized by three
scalar form factors \cite{Pagels,Ji}. These form factors describe
the mechanical structure of the spin$-\frac 12$ system--nucleon.
At zero momentum transfer these three form factors can be related to
three basic static mechanical characteristics of  the composite
system -- nucleon. Two of them are
energy-momentum and angular momentum carried by partons. The third
one is related to the constants $d_1(0)$ and $d^G_1(0)$ in
expansion of the D-term, see eqs.~(\ref{dterm},\ref{dtermg}).
These constants can be interpreted as the characteristic of the
spatial distribution of the stress tensor or as of the
spatial distribution of ``forces" experienced by various species
of partons inside of the nucleon.
At $t=0$ we can express the $x$-moments of the GPDs as follows
\cite{Ji}:
\be
\label{jisr}
\int_{-1}^1 dx\ x\ \left( H(x,\xi)+E(x,\xi)\right)=2J^Q\, ,\\
\int_{-1}^1 dx\ x\ H(x,\xi)=M_2^Q +\frac 45\ d_1\ \xi^2\, .
\nonumber
\ee
Here $J^Q$ is a fraction of the nucleon angular momentum carried
by all quarks, $d_1$ is the first Gegenbauer coefficient in the expansion of
the D-term (\ref{dterm}) and $M_2^Q$ is a momentum fraction carried
by quarks and antiquarks in the nucleon:
\be
M_2^Q= \sum_q \int_0^1dx\ x \ \left(q(x) + \bar q(x) \right)\, .
\ee
Analogous expressions one can write for gluon GPDs. To see physics
interpretation of nucleon ``mechanical"  characteristics $J^Q,M_2^Q$ and
$d_1$ let us write their relations to the static energy-momentum
tensor of the nucleon. In the Breit frame (in which the energy transfer $\Delta^0=0$)
and non-relativistic limit ($\vec \Delta^2 \ll m_N^2$) we can define static
energy-momentum tensor of the nucleon as (see e.g. \cite{Donoghue:2001qc}):

\be
T_{\mu\nu}^Q(\vec r)=\frac{1}{2m_N}\int \frac{d^3\Delta}{(2\pi)^3}\
e^{i\vec{r}\cdot\vec{\Delta}}\
\langle p',S| \hat T_{\mu\nu}^Q(0)|p,S\rangle\, ,
\ee
where $\hat T_{\mu\nu}^Q(0)$ is the QCD operator of the symmetric energy
momentum tensor of quarks\footnote{An analogous equation and the discussion below can be
trivially extended for gluons as well} and $S$ is the polarization vector of the nucleon.
 Now it is easy to express the
constants $J^Q,M_2^Q$ and $d_1$ in the following way:

\be
\nonumber
M_2^Q&=&\frac{1}{m_N}\ \int d^3 r\ T_{00}^Q(\vec r)\\
\nonumber
J^Q&=& \int d^3 r\ \varepsilon^{i j k}\ S_i\ r_j\ T_{0k}^Q(\vec r)\, .
\ee
These relations illustrate the interpretation of $M_2^Q$ and $J^Q$
as fractions of energy-momentum and angular momentum of the
nucleon carried by quarks and antiquarks. This follows from the
fact that $T_{00}^Q(\vec r)$ and $T_{0i}^Q(\vec r)$ can be
interpreted as a spatial distribution of energy and momentum carried by quarks
correspondingly.

The constant $d_1$ can
be expressed in terms of the stress tensor $T_{ij}(\vec r)$:

\be
\label{EMTd1}
d_1=-\frac{m_N}{2}\  \int d^3r\ \ T_{ij}^Q(\vec r)\ \left(r^i r^j-\frac
13\
\delta^{ij} r^2\right)\, .
\ee
If one would consider the nucleon as a continuous medium than $T_{ij}^Q(\vec r)$
would characterize the force experienced by quarks in an infinitesimal volume
at distance $\vec r$ from the centre of the nucleon. More detailed
distribution of forces one can obtain from the $t$-dependence of
$d_1(t)$ because of relation

\be
\int \frac{d^3 \Delta}{(2\pi)^3}\ e^{-i\vec{r}\cdot\vec{\Delta}} d_1\left(-{\vec \Delta}^2\right)\propto
T_{ij}^Q(\vec r)\ \left(r^i r^j-\frac
13\
\delta^{ij} r^2\right)\, .
\ee

Obviously the sums $M_2^Q+M_2^G=1$, $J^Q+J^G=1/2$ and $d_1+d_1^G=d$
are scale independent. Two first sums corresponds to the total momentum and total
angular momentum of the nucleon. The third constant $d$ using eq.~(\ref{EMTd1}) for
the \underline{conserved} total energy-momentum tensor can be
represented in the form:

\be
d=\frac{5m_N}{9}\ \int d^3 r\ r^2 \ p(r)\, ,
\ee
where we introduced the following parametrization of the total (quarks+gluons)
static stress tensor:

\be
T_{ij}(\vec r)=s(r) \frac{r_ir_j}{r^2}+p(r)\delta_{ij}\, .
\ee
The function $s(r)$ and $p(r)$ are related to each other  by conservation of the
total energy-momentum tensor. The function $p(r)$ can be
interpreted as the radial distribution of the ``pressure" inside the nucleon.
We note also that the $t$-dependence of the GPDs provides us with more detailed
spatial images
of the nucleon, see e.g. \cite{Bur,Ral,Die}.

The estimate
which is based on the calculation of GPDs
in the chiral quark soliton model \cite{Pet98}
at a low normalization point $\mu\approx 0.6$~GeV,
gives \cite{Kiv} rather large and negative value of $d_1 \approx
-4.0$. The negative values of this constant has a deep relation to the spontaneous breaking of the chiral symmetry
in QCD, see \cite{MVP98,Kiv,GPV}.

\section*{\normalsize\bf Partial wave decomposition of the GPDs}

Here we discuss the decomposition of GPDs in $t$-channel partial
waves. Such kind of decomposition is useful for
understanding of physical mechanism contributing to
generalized parton distributions. In a sense we attempt to model
GPDs not by ``deforming" (``skewing") the forward parton
distributions but rather representing the GPDs as an infinite sum
of the distribution amplitudes in the $t$-channel. Using the
analytical continuation of the $t$-channel exchange representation we shall try to
relate two ``dual" approaches to the GPDs.

The partial wave decomposition in the $t$-channel for singlet GPDs $H(x,\xi,t)$
and $E(x,\xi,t)$ can be written as the following formal series \cite{MVP98}:

\be
\label{formal}
H(x,\xi,t)&=&
\sum_{n=1\atop \scriptstyle{\rm odd}}^\infty \sum_{l=0\atop \scriptstyle{\rm even}}^{n+1}
B_{nl}(t)\
\theta\left(1-\frac{x^2}{\xi^2}\right)\left(1-\frac{x^2}{\xi^2}\right)
\ C_n^{3/2}\left(\frac{x}{\xi}\right) \ P_l\left(\frac 1
\xi\right)\, ,\\
E(x,\xi,t)&=&
\sum_{n=1\atop \scriptstyle{\rm odd}}^\infty \sum_{l=0\atop \scriptstyle{\rm even}}^{n+1} C_{nl}(t)\
\theta\left(1-\frac{x^2}{\xi^2}\right)\left(1-\frac{x^2}{\xi^2}\right)
\ C_n^{3/2}\left(\frac{x}{\xi}\right) \ P_l\left(\frac 1
\xi\right)\, .
\ee
Here the index $l$ corresponds to
an exchange in the $t$--channel with the orbital momentum $l$, because in the hard regime
the $t$-channel scattering angle $\theta_t$ is related to $\xi$ via $\cos\theta_t=1/\xi$.
The coefficients $B_{nl}(t)$ and $C_{nl}(t)$ are generalized form factors. Note
that for the fixed power of the Gegenbauer polynomial $n$ the
orbital momentum can reach maximally the value $l=n+1$, this is the
consequence of the Lorentz invariance \cite{MVP98}. The formal series (\ref{formal}) corresponds to
analytical continuation (crossing) of the corresponding expansion for the generalized distribution amplitudes
\cite{DiePirTer}
entering the description of hard processes like $\gamma^*\gamma\to h\bar h$,
see discussion in refs.~\cite{MVP98,OlegRadon}.

Below we shall
discuss mostly the GPD $H(x,\xi,t)$ the expressions for $E(x,\xi,t)$
are obtained by a trivial generalization.
The
form factors $B_{n n+1}(t=0)$ at zero momentum transfer
are fixed in terms of the Mellin moments of usual singlet quark
distributions \cite{MVP98}:
\be
\label{crossing}
B_{n n+1}(t=0)=\ \frac{2 n+3}{2(n+2)}\int_0^1 dx\ x^n
\left( q(x)+\bar q(x)\right)\, .
\ee
The other coefficients $B_{nl}$ with $l\leq n-1$ are not related
to the forward distribution and thus are ``truly non-forward"
quantities.

Each term in the sum (\ref{formal}) has a support $-\xi < x<\xi$,
however this does not imply that the GPD
$H(x,\xi,t)$ has the same support because generically the sum (\ref{formal})
is divergent at fixed $x$ and $\xi$. In the formal solution
(\ref{formal}) each individual term of the series gives the
contribution to the amplitude which behaves as
$1/\xi^l$ at small $\xi$ what would imply the violation of
unitarity for the hard exclusive reactions. This ``disaster" is avoided
through the fact that
the series is divergent and one can not calculate the
asymptotic behaviour term by term. This situation is similar to
duality idea: the $entire$ generalized
and usual quark distributions  are given by infinite sum over
$t$--channel exchanges.

The formal representation (\ref{formal})
can be equivalently rewritten as the convergent series\footnote{
 Actually this formula can be effectively used
for numerical construction of GPDs from the coefficients $B_{nl}$ }:
\be
\label{convergent}
H(x,\xi,t)=(1-x^2) \sum_{n=1\atop \scriptstyle{\rm odd}}^\infty A_n(\xi,t)\
C_n^{3/2}(x)\, ,
\ee
where
\be
A_n(\xi,t)=-\frac{2 n+3}{(n+1)(n+2)}\ \sum_{p=1\atop \scriptstyle{\rm odd}}^n\xi\ R_{np}(\xi)\
\frac{(p+1)(p+2)}{2p+3}\ \sum_{l=0\atop \scriptstyle{\rm even}}^{p+1} B_{pl}(t)\ P_l\left(\frac
1\xi\right) \, .
\ee
Here $R_{np}(\xi)$ are polynomials in $\xi$ of the order $n$
introduced in ref.~\cite{Belitsky:1998pc,Kiv99}:

\be
R_{np}(\xi)=
(-1)^{
\frac{n + p}{2}}
\frac{\Gamma(\frac{3}{2} + \frac{p + n}{2})}{
\Gamma(\frac{n - p}{2}+1) \Gamma(\frac 32 + p)} \xi^p\
{}_2F_1
\left(\frac p2 - \frac n2, \frac 32 + \frac n2 + \frac p2;
\frac 52 + p; \xi^2\right)\, .
\ee
In the series (\ref{convergent}) at small $\xi$ the terms with
$l=n+1$ are dominant, so that in the limit $\xi\to 0$ the forward distribution is recovered from
eq.~(\ref{convergent}). Corrections to the forward limit of order $\xi^2$ at fixed $x$ is completely determined by
orbital momenta of $l=n+1$ and $l=n-1$. We see that the smaller skewedness $\xi$ the less important the deviations of the
the orbital momentum $l$ from its maximal value of $n+1$. We shall
use this fact in next section to perform summation over orbital
momenta.

{}From either (\ref{formal}) or (\ref{convergent}) we can easily
obtain the partial wave decomposition of the Mellin moments \cite{MVP98}:

\be
\label{pwamoments}
\int_{-1}^1 dx\ x^N\ H(x,\xi,t)=\xi^{N+1}
\sum_{n=1\atop \scriptstyle{\rm odd}}^{N}\sum_{l=0\atop \scriptstyle{\rm even}}^{n+1}
B_{n l}(t)\ P_l\left(\frac 1\xi\right)
\frac{\Gamma\left(\frac 32\right)\Gamma(N+1) (n+1)(n+2)}{2^n
\Gamma\left(\frac{N-n}{2}+1\right)\Gamma\left(\frac{N+n}{2}+\frac
52\right)}\, ,
\ee
which obviously satisfies the polynomiality condition (\ref{pc}).
For the lowest $N=1$ moment we have:
\be
\int_{-1}^1 dx\ x\ H(x,\xi,t)&=&\frac{6}{5} \left[
B_{12}(t)-\frac 13 (B_{12}(t)-2 B_{10}(t))\ \xi^2 \right]\, ,\\
\int_{-1}^1 dx\ x\ E(x,\xi,t)&=&\frac{6}{5} \left[
C_{12}(t)-\frac 13 (C_{12}(t)-2 C_{10}(t))\ \xi^2 \right]\, ,
\ee
which with help of relations (\ref{crossing})
and (\ref{jisr}) gives at $t=0$:
\be
\int_{-1}^1 dx\ x\ H(x,\xi)&=&
M_2^Q-\frac 13 \left(M_2^Q-\frac{12}{5}\ B_{10}(0)\right)\ \xi^2\,
,\\
\int_{-1}^1 dx\ x\ E(x,\xi)&=&
2 J^Q-M_2^Q-\frac 13 \left(2J^Q-M_2^Q-\frac{12}{5}\ C_{10}(0)\right)\ \xi^2\, .
\ee
{}From this expression we can easily obtain the value of the
first Gegenbauer coefficient of the D-term $d_1$ at $t=0$:
\be
\label{d1m2}
d_1 = -\frac{5}{12} M_2^Q + B_{10}(0) \, .
\ee
Equivalently we can extract the $d_1$ using the fact that the
D-term is the same (up to the sign) for $H$ and $E$:
\be
\label{d1j}
d_1 =\frac{5}{12}\left( 2J^Q- M_2^Q\right) - C_{10}(0) \, .
\ee
Comparing eq.~(\ref{d1m2}) with (\ref{d1j}) we obtain new
representation for
the Ji's sum rule in terms of S-wave exchanges:
\be
\label{newjisr}
J^Q=\frac 65 \left(B_{10}(0)+C_{10}(0)\right)\, ,
\ee
which should be contrasted with the original formulation of this
sum rule done in terms of D-wave exchanges:
\be
J^Q=\frac 35 \left(B_{12}(0)+C_{12}(0)\right)\, .
\ee
We see that the $J^Q$ can be extracted from the S-wave exchange in
the t-channel. This is not surprising because between functions
$H(x,\xi)$ and $E(x,\xi)$ there are nontrivial functional
relations, see e.g. eq.~(\ref{crit}).
Generically for the $N$--th Gegenbauer coefficient  of the D-term
we have two equivalent representations
\be
d_N(t)&=&\sum_{l=0\atop \scriptstyle{\rm even}}^{N+1} B_{Nl}(t)\ \frac{(-1)^{l/2} \Gamma\left(\frac l2+\frac 12\right)}{
\Gamma\left(\frac l2+1\right)\Gamma\left(\frac 12\right)}\, ,\\
d_N(t)&=&-\sum_{l=0\atop \scriptstyle{\rm even}}^{N+1} C_{Nl}(t)\ \frac{(-1)^{l/2} \Gamma\left(\frac l2+\frac 12\right)}{
\Gamma\left(\frac l2+1\right)\Gamma\left(\frac 12\right)}\, .
\ee
This implies the following relations between coefficients
$B_{Nl}(t)$ and $C_{Nl}(t)$:
\be
\label{bvsc}
\sum_{l=0\atop \scriptstyle{\rm even}}^{N+1} B_{Nl}(t)\ \frac{(-1)^{l/2} \Gamma\left(\frac l2+\frac 12\right)}{
\Gamma\left(\frac l2+1\right)\Gamma\left(\frac 12\right)}=
-\sum_{l=0\atop \scriptstyle{\rm even}}^{N+1} C_{Nl}(t)\ \frac{(-1)^{l/2} \Gamma\left(\frac l2+\frac 12\right)}{
\Gamma\left(\frac l2+1\right)\Gamma\left(\frac 12\right)}\, .
\ee
Such kind of relations is useful for modeling of GPDs.

The expression of the $J^Q$ in terms of S-wave t-channel exchanges
(\ref{newjisr})
is useful because it allows us to analyze $J^Q$ using various
models and methods of low energy physics.

\section*{\normalsize\bf Summing up the partial waves}

The partial wave decomposition of GPDs discussed in previous
section can be used to write a parametrization of the GPDs in
terms of ``forward like" functions. To do this we introduce a set
of functions whose Mellin moments generate the coefficients
$B_{nl}(t)$ and
$C_{nl}(t)$
\be
B_{n n+1-k}(t)=
 \int_0^1 dx\ x^n\ Q_k(x,t)\, ,\\
C_{n n+1-k}(t)=
 \int_0^1 dx\ x^n\ R_k(x,t)\, ,
\label{def-Q}
\ee
where index $k$ (always even number!) characterizes the deviation
of the orbital momentum from the maximally possible value of $l=n+1$.
Note that the LO evolution of the functions $Q_k(x,t)$ and $R_k(x,t)$ is
governed by the DGLAP evolution equation.
Since $B_{nn+1}(0)$ is fixed by the Mellin moments of forward
distributions, see eq.~(\ref{crossing}), the function $Q_0(x,t=0)$
is related to forward distributions:

\be
Q_0(x)=\left[ (q(x)+\bar q(x))-\frac x2  \int_x^1 \frac{dz}{z^2}\ ( q(z)+\bar q(z))
\right]\, .
\label{Q0}
\ee

We sequence of functions\footnote{The discussion below is applied
to the
functions $R_k(x,t)$ as well} $Q_0(x,t), Q_2(x,t), Q_4(x,t), \ldots$ is
introduced in such a way that the higher functions (with high index $k$) are more and
more suppressed for small skewedness parameter $\xi$.

Using the
methods developed in refs.~\cite{Shu,Andrei} we can derive the
following integral transformation relating GPDs $H(x,\xi,t)$\footnote{Remember that
here we discuss only the singlet ($C=+1$) GPDs, collection of formulas for nonsinglet
($C=-1$) GPDs
is given in the Appendix~A} and
functions $Q_k(y,t)$:

\be
\label{integralrelation}
H(x,\xi,t)&=&\sum_{k=0\atop \scriptstyle{\rm even}}^{\infty}\Biggl\{ \frac{\xi^k}{2}\
\left[ H^{(k)}(x,\xi,t)-H^{(k)}(-x,\xi,t)\right]\\
\nonumber
&+&
\left(1-\frac{x^2}{\xi^2}\right)\theta(\xi-|x|)\sum_{l=1\atop \scriptstyle{\rm odd}}^{k-3}
C^{3/2}_{k-l-2}\left(\frac{x}{\xi}\right)\ P_l\left(\frac
1\xi\right)\int_{0}^1 dy\ y^{k-l-2}\ Q_k(y,t)\Biggr\}\, ,
\ee
where the function $H^{(k)}(x,\xi,t)$ is given as the following
integral transformation:

\be
\nonumber
H^{(k)}(x,\xi,t)&=&\theta\left(x>\xi\right)\ \frac{1}{\pi}
\int_{y_0}^1 \frac{dy}{y}\left[\left(1-y\frac{\partial}{\partial y}\right)
Q_k(y,t)\right]
\int_{s_1}^{s_2} ds \frac{x_s^{1-k}}{\sqrt{x_s^2-2x_s-\xi^2}}\\
&+&
\label{kernel-podrobno}
\theta\left(x<\xi\right)\ \frac{1}{\pi}
\int_{0}^1 \frac{dy}{y}\left[\left(1-y\frac{\partial}{\partial y}\right)
Q_k(y,t)\right]
\int_{s_1}^{s_3} ds \frac{x_s^{1-k}}{\sqrt{x_s^2-2x_s-\xi^2}}\, .
\ee
Here $x_s=2\frac{x-s\xi}{(1-s^2)y}$ and integration limits $s_1,s_2, s_3$ and $y_0$ are
given by the following expressions:

\be
\nonumber
s_1&=&\frac{1}{y\xi}\left[1-\sqrt{1-\xi^2}-\sqrt{2\left(1-x
y\right)\left(1-\sqrt{1-\xi^2}\right)-\xi^2\left(1-y^2\right)}\ \right]\, ,\\
\nonumber
s_2&=&\frac{1}{y\xi}\left[1-\sqrt{1-\xi^2}+\sqrt{2\left(1-x
y\right)\left(1-\sqrt{1-\xi^2}\right)-\xi^2\left(1-y^2\right)}\ \right]\, ,\\
\nonumber
s_3&=&\frac{1}{y\xi}\left[1+\sqrt{1-\xi^2}-\sqrt{2\left(1-x
y\right)\left(1+\sqrt{1-\xi^2}\right)-\xi^2\left(1-y^2\right)}\ \right]\, ,\\
y_0&=&\frac{1}{\xi^2}\left[x \left(1-\sqrt{1-\xi^2}\right)+
\sqrt{ \left(x^2-\xi^2\right)\left(2\left(1-\sqrt{1-\xi^2}\right)-\xi^2\right)}\ \right]\, .
\ee
Actually the integral transformation (\ref{kernel-podrobno}) can
be written compactly as:

\be
\label{kernel}
H^{(k)}(x,\xi,t)=\frac{1}{\pi}
\int_{0}^1 \frac{dy}{y}\left[\left(1-y\frac{\partial}{\partial y}\right)
Q_k(y,t)\right]
\int ds \frac{x_s^{1-k}}{\sqrt{x_s^2-2x_s-\xi^2}}\
\theta(x_s^2-2x_s-\xi^2)\, .
\ee
Derivation of eqs.~(\ref{kernel-podrobno},\ref{kernel}) is given
in the Appendix~B. Also we note that the numerical realization of these equations is stable and fast,
it takes fraction of a second on a PC to compute GPDs  for given function $Q_k(x,t)$.

The simple analytical application of these integral transformations is to estimate the
small$-\xi$, Regge-type, behaviour of the imaginary part of the leading order
amplitude proportional to $\sum_k H^{(k)}(\xi,\xi)$.
Taking the power-like parametrization for the
function $Q_k(x)$ in the region of small $y$--$Q_k(y) \sim y^{-\alpha}$,
we get (for $\alpha>k-1/2$)\footnote{For the case of $\alpha < k-1/2$ we
obtain that $
H^{(k)}(\xi,\xi)\,\sim\ \sqrt{\xi}$ with the coefficient which depends also
on behaviour of the function $Q_k(y)$ at large $y$.}:
\be
\label{smallxi}
\xi^k\ H^{(k)}(\xi,\xi)\,\sim\,\frac 1\pi\, \left(\frac{\xi}{2}\right)^{k-\alpha}\,
\frac {\Gamma(1/2)\Gamma(\alpha-k+1/2)}{\Gamma(\alpha-k+1)},
\ee
which is in an agreement with the result of Ref.~\cite{Shu,Andrei}
for $k=0$. Also we see that for the case when the $\alpha$ does
not increase strongly with increasing of $k$ the leading term is
determined by the functions $Q_0(x)$ which is completely fixed by
the forward distributions.

The original idea of Refs.~\cite{Andrei,Shu} to relate the GPDs to functions which are evolved
according to usual DGLAP equation, as pointed out in Refs.~\cite{Nor,Mus}, implies
nontrivial ($\xi$ dependent) constrains for the support of the
``effective forward functions". This makes the parametrization of these functions almost impractical.
Our construction, given by eq.~(\ref{kernel-podrobno}), seems to be free of this problem.

The leading order amplitude of hard exclusive reactions is expressed
in terms of the following elementary amplitude\footnote{We restrict ourselves to the
singlet or even signature amplitudes only. For the nonsinglet (odd signature) amplitude
see Appendix~A.}:
\be
A(\xi,t)=\int_{0}^1 dx\,H(x,\xi,t)\,
\left[\frac 1{x-\xi+i0} +\frac 1{x+\xi-i0}\right].
\label{elementaryAMP}
\ee
Now we can express the amplitudes in terms of ``forward-like" functions
$Q_k(x)$. For this we substitute the formal series representation
for GPDs (\ref{formal}) into eq.~(\ref{elementaryAMP}) and obtain
the partial wave decomposition of the amplitude in the
$t$-channel:

\be
A(\xi,t)=-2\sum_{n=1\atop \scriptstyle{\rm odd}}^\infty \sum_{l=0\atop \scriptstyle{\rm even}}^{n+1}
B_{nl}(t) \ P_l\left(\frac 1
\xi\right)\, .
\ee
Substituting into this equation the expression for the
coefficients $B_{nl}(t)$ in terms of functions $Q_k(x,t)$
(\ref{def-Q}), we can sum up the partial waves with the result:

\begin{eqnarray}
A(\xi,t)&=&
-\int_0^1 \frac{dx}{x}\sum_{k=0}^{\infty} x^k Q_k(x,t)
\Biggl[
\frac{1}{\sqrt{1-\frac{2 x}{\xi}+x^2}} +
\frac{1}{\sqrt{1+\frac{2 x}{\xi}+x^2}}-2\ \delta_{k0}
\Biggr]\,
\end{eqnarray}
Using this relation we can write down explicit expressions
for the real and imaginary parts of the amplitudes:

\begin{eqnarray}
\nonumber
{\rm Im\ } A(\xi,t)&=&
-\int_{\frac{1-\sqrt{1-\xi^2}}{\xi}}^1
\frac{dx}{x}\sum_{k=0}^{\infty} x^k Q_k(x,t)
\Biggl[
\frac{1}{\sqrt{\frac{2 x}{\xi}-x^2-1}}
\Biggr]\, ,\\
\nonumber
{\rm Re\ } A(\xi,t)&=&
-\int_0^{\frac{1-\sqrt{1-\xi^2}}{\xi}}
\frac{dx}{x}\sum_{k=0}^{\infty} x^k Q_k(x,t)
\Biggl[
\frac{1}{\sqrt{1-\frac{2 x}{\xi}+x^2}} +
\frac{1}{\sqrt{1+\frac{2 x}{\xi}+x^2}}-2\ \delta_{k0}
\Biggr]  \\
&-&\int^1_{\frac{1-\sqrt{1-\xi^2}}{\xi}}
\frac{dx}{x}\sum_{k=0}^{\infty} x^k Q_k(x,t)
\Biggl[
\frac{1}{\sqrt{1+\frac{2 x}{\xi}+x^2}}-2\ \delta_{k0}
\Biggr]
\, .
\label{REIM}
\end{eqnarray}
These expressions would allow us to analyze the experimental data
on hard exclusive reactions in terms of ``forward-like" functions
$Q_k(x,t)$ and $R_k(x,t)$. Among these functions only $Q_0(x)$ is fixed in terms of
forward distributions, see eq.~(\ref{Q0}). Note also that the
smaller $\xi$ the more suppressed the contribution of the functions
$Q_k(x,t)$ and $R_k(x,t)$ with higher $k$ to the amplitude.
We also note that from point of view of practical applications the numerical calculations of the
amplitude according our formulae (\ref{REIM}) is more stable and faster than calculations of more
singular integral (\ref{elementaryAMP}). From expression (\ref{REIM}) for the imaginary part of the amplitude
(corresponding to $-\pi H(\xi,\xi,t)$) we see that for its calculation we need to know the functions $Q_k(x)$
(and hence forward parton distributions) for $x\geq (1-\sqrt{1-\xi^2})/\xi>\xi/2$. In this way the problem (discussed in
refs.~\cite{FreMcD})
that in the
parametrization of GPDs in terms of double distributions one needs to sample forward distributions down to very
small (unmeasured) values of $x$ is solved. In our parametrization we see explicitly that the contribution to the amplitude
of very small $x$ is strongly suppressed as it should be from physics point of view.

{}From the expressions (\ref{REIM}) one can easily obtain the behaviour
of the amplitude at small $\xi$. Indeed, let us assume that at small $x$ we have
$Q_k(x,t)\sim x^{-\alpha_k}$, then the contribution of each individual function $Q_k(x,t)$
to the amplitude has the form:

\be
\label{imsmall}
{\rm Im} A^{(k)}(\xi,t)\sim -\left(\frac{\xi}{2}\right)^{k-\alpha_k}\,
\frac {\Gamma(1/2)\Gamma(\alpha_k-k+1/2)}{\Gamma(\alpha_k-k+1)}\, \\
{\rm Re} A^{(k)}(\xi,t)\sim {\rm Im} A^{(k)}(\xi,t)\
\tan\left(\frac{\pi (\alpha_k-k-1)}{2}\right)\,.
\label{resmall}
\ee
The first equation coincides with the small $\xi$ behaviour of
$-\pi H^{(k)}(\xi,\xi)$ (see eq.~(\ref{smallxi}))
as it should be. The second equation reproduces the result of general
dispersion relations for the even signature scattering amplitudes
at high energies \cite{Gribov}. Also we see that if $\alpha_k$ does not grow
drastically with $k$, the
leading contribution comes from the function $Q_0(x)$ which is
fixed in terms of the forward parton distributions.

To derive eqs.~(\ref{imsmall},\ref{resmall}) we note that at small
values of the skewedness parameter $\xi$ the expressions for the
real and imaginary part of the amplitude (\ref{REIM}) can be
rewritten for the case $\alpha_k-k>1/2$ as
(after rescaling of the integration variables $2 x/\xi\to x$):

\be
{\rm Im\ } A(\xi,t)&\sim& -\sum_{k=0}^{\infty}
\left(\frac{\xi}{2}\right)^{k-\alpha_k}\int_1^\infty
\frac{dx}{x^{\alpha_k-k+1}}\left[\frac{1}{\sqrt{x-1}}\right]\, ,
\label{imsm}\\
\nonumber
{\rm Re\ } A(\xi,t)&\sim& -\sum_{k=0}^{\infty}
\left(\frac{\xi}{2}\right)^{k-\alpha_k}\int_0^1
\frac{dx}{x^{\alpha_k-k+1}}\left[\frac{1}{\sqrt{1-x}}+\frac{1}{\sqrt{1+x}}-2\delta_{k0}\right]\\
&-&\int_1^\infty\frac{dx}{x^{\alpha_k-k+1}}
\left[\frac{1}{\sqrt{1+x}}-2\delta_{k0}\right]\, .
\label{resm}
\ee
Performing integrals in the above expressions we obviously obtain
eqs.~(\ref{imsmall},\ref{resmall}).

Using eqs.~(\ref{REIM}) one can try to fit experimental data on
hard exclusive processes adopting a simple parametrization of the
functions $Q_k(x,t)$ (for $k\geq 2$) borrowed from analysis of the DIS data:

\be
Q_k(x,t)=N_k\ \frac{1}{x^{\alpha_k}}\
\left(1-x\right)^{\beta_k}\left(1+\gamma_k\sqrt{x} +\delta_k\ x
\right)\, ,
\ee
and analogous form for the functions $R_k(x)$. Where, in principle, all parameters $\alpha_k, \beta_k$, etc.
are functions of the momentum transfer squared $t$\footnote{The minimal possible choice of the $t$-dependence
can be $\alpha_k(t)=\alpha_k +\alpha'\ t$ \cite{GPV} where $\alpha'$ is the slope of the corresponding Regge trajectory.}.
Introduced
parameters for $k=0, 2$ at $t=0$ can be related to the basic mechanical characteristics of the nucleon--$M_2^Q$, $J^Q$ and
$d_1$ by:

\be
\nonumber
\int_0^1dx\ x\ Q_0(x)&=&\frac{5}{6}\ M_2^Q\, , \\
\nonumber
\int_0^1dx\ x\ Q_2(x)&=&d_1+\frac{5}{12}\ M_2^Q\, , \\
\label{pravilasum}
\int_0^1dx\ x\ R_0(x)&=&\frac{5}{6}\left(2J^Q- M_2^Q\right)\, ,\\
\int_0^1dx\ x\ R_2(x)&=&\frac{5}{12}\left(2J^Q- M_2^Q\right)-d_1\, .
\nonumber
\ee
Note also that the leading order evolution of the functions $Q_k(x,t)$ and $R_k(x,t)$ is     given by the
usual DGLAP equations, so that for their evolution one can use fast and effective algorithms developed for the DIS,
see e.g. Refs.~\cite{evol}. Further constrains on the parameters $\alpha_k, \beta_k$, etc. can be obtained
from the positivity bounds for GPDs \cite{Terpos,Radpos,Diehlpos}
derived in the most complete form recently by Pobylitsa \cite{Pasha}.

\section*{\normalsize\bf ``Minimal model"}
Discussions in previous sections give us a hope that in the analysis of the data on hard exclusive processes
we can restrict ourselves to the finite set of the ``forward-like" functions $Q_k(x,t)$ and $R_k(x,t)$. This
set can be systematically extended with increasing the amount of experimental data. The most interesting
``mechanical" properties of the nucleon are contained in the lowest functions for $k=0,2$.
In this section we discuss possible minimal set of the ``forward-like" functions. Clearly the choice with only
$k=0$ is too restrictive, in particular, such a choice would imply that $H(x,\xi,t)=-E(x,\xi,t)$ which is too
strong constrain.
One may try
to fit the data with functions $Q_0(x,t)$ and $Q_2(x,t)$ (and $R_0(x,t)$, $R_2(x,t)$) one of
those is fixed by the forward parton distributions.
In the model with only two types of generating functions
$Q_0(x,t)$ and $Q_2(x,t)$ (and $R_0(x,t)$, $R_2(x,t)$) one can easily
derive from the constrain (\ref{bvsc}) the functional relation for the
function $R_2(x,t)$:
\be
\label{r2}
R_2(x,t)=-Q_2(x,t)+Q_0(x,t)+R_0(x,t)-\int_x^1 \frac{dz}{z}\
\left[Q_0(z,t)+R_0(z,t)\right]\, .
\ee
We see that in such model one has only two new ``forward-like" functions ($Q_2(x,t)$ and
$R_0(x,t)$),
$Q_0(x,t)$ is essentially fixed by the forward distributions, see eq.~(\ref{Q0})\footnote{
An analogous
expression can be also written
for $R_0(x)$ where in rhs the role of forward distribution
is played by $E(x,\xi=0)$. }, and
$R_2(x,t)$ by the relation (\ref{r2}).

In this paper we do not present our numerical results for GPDs in
the ``minimal model" just discussed, the
corresponding results we shall present elsewhere.
Our main aim here was to set the theoretical framework for
modeling of GPDs.

\section*{\normalsize\bf Conclusions and outlook}
We analyzed the representation of the generalized parton distributions (GPDs) as the infinite sum
over distribution amplitudes in the $t$-channel. Such
representation is close in its spirit to the duality idea--the
entire {\em parton distribution} is obtained as an infinite sum over
{\em distribution amplitudes} in the $t$-channel.
On basis of this dual representation we suggested a parametrization
of the GPDs which satisfy automatically all general constrains-
forward limit, polynomiality, etc.
In this parametrization the GPDs and the leading order amplitudes are
expressed in terms of ``forward-like"
functions $Q_k(x,t)$ ($R_k(x,t)$). The dependence of these
functions on the scale
in the leading order is governed by
the DGLAP evolution equation.
The index $k$ which enumerates the set of functions $Q_k(x,t)$ ($R_k(x,t)$)
describes the deviation of the orbital momentum
$l$ in the $t$-channel from its maximal value of $l=n+1$
at fixed conformal spin $n$ of the twist-2 operator. The basic
mechanical characteristics of the nucleon--momentum and angular
momentum fractions carried by quarks and additionally the
radial distribution of forces experienced by quarks in the
nucleon--are contained in the lowest functions $Q_0(x,t),Q_2(x,t)$ ($R_0(x,t),R_2(x,t)$).

With our parametrization of GPDs still there are several points which
should be worked out further:
\begin{itemize}
\item
To prove that suggested parametrization covers all possible forms of GPDs and derive the inversion
formula expressing functions $Q_k$ and $R_k$ through GPDs $H$ and $E$.
\item
To extend the analysis to the next to leading order.
On NLO evolution of GPDs see, e.g. ref.~\cite{BelMul}.
\end{itemize}

\section*{\normalsize\bf Acknowledgements}
We are thankful to B.~Holstein, I.B.~Khriplovich, N.~Kivel, P.~Pobylitsa and
M.~ Vanderhaeghen for many
valuable discussions. The
work is supported
by the Sofja Kovalevskaja Programme of the Alexander von Humboldt
Foundation, the Federal Ministry of Education and Research and the
Programme for Investment in the Future of German Government.

\section*{\normalsize\bf Appendix A}
In the main text we discussed only the singlet ($C=+1$) GPDs which
are reduced to $[q(x)+\bar q(x)]$ in the forward limit. In this
Appendix we collect formulae for the nonsiglet ($C=-1$) GPDs which
in the forward limit correspond to $[q(x)-\bar q(x)]$. Below we
simply list the corresponding formulae, skipping the discussion.\\
\underline{Partial wave decomposition}:

\be
\label{formalnn}
H(x,\xi,t)&=&
\sum_{n=0\atop \scriptstyle{\rm even}}^\infty \sum_{l=1\atop \scriptstyle{\rm odd}}^{n+1}
B_{nl}(t)\
\theta\left(1-\frac{x^2}{\xi^2}\right)\left(1-\frac{x^2}{\xi^2}\right)
\ C_n^{3/2}\left(\frac{x}{\xi}\right) \ P_l\left(\frac 1
\xi\right)\, ,\\
E(x,\xi,t)&=&
\sum_{n=0\atop \scriptstyle{\rm even}}^\infty
\sum_{l=1\atop \scriptstyle{\rm odd}}^{n+1} C_{nl}(t)\
\theta\left(1-\frac{x^2}{\xi^2}\right)\left(1-\frac{x^2}{\xi^2}\right)
\ C_n^{3/2}\left(\frac{x}{\xi}\right) \ P_l\left(\frac 1
\xi\right)\, .
\ee
\underline{Mellin moments}:
\be
\label{pwamomentsnn}
\int_{-1}^1 dx\ x^N\ H(x,\xi,t)=\xi^{N+1}
\sum_{n=1\atop \scriptstyle{\rm even}}^{N}\sum_{l=0\atop \scriptstyle{\rm odd}}^{n+1}
B_{n l}(t)\ P_l\left(\frac 1\xi\right)
\frac{\Gamma\left(\frac 32\right)\Gamma(N+1) (n+1)(n+2)}{2^n
\Gamma\left(\frac{N-n}{2}+1\right)\Gamma\left(\frac{N+n}{2}+\frac
52\right)}\, .
\ee
The lowest $x^0$ Mellin moments of GPDs are related to the
electromagnetic form factors of the nucleon

$$
\int_{-1}^1 dx\ H(x,\xi,t)=F_1(t)=\frac{2}{3} B_{01}(t)\, ,
$$
and
$$
\int_{-1}^1 dx\ E(x,\xi,t)=F_2(t)=\frac{2}{3} C_{01}(t)\, ,
$$
\underline{Generating functions $Q_k$ and $R_k$} are introduced in
the same way as in eqs.~(\ref{def-Q}) only in the nonsinglet case the number $n$ is
even. The function $Q_0(x,t=0)$ is related to the nonsinglet
forward quark distribution:

\be
Q_0(x)= \left[ (q(x)-\bar q(x))-\frac x2  \int_x^1 \frac{dz}{z^2}\ ( q(z)-\bar q(z))
\right]\, .
\label{Q0nn}
\ee
\underline{Integral transformations $Q_k\to H$.}
\be
H(x,\xi,t)&=&\sum_{k=0\atop \scriptstyle{\rm even}}^{\infty}\Biggl\{ \frac{\xi^k}{2}\
\left[ H^{(k)}(x,\xi,t)+H^{(k)}(-x,\xi,t)\right]\\
\nonumber
&+&
\left(1-\frac{x^2}{\xi^2}\right)\theta(|x|-\xi)\sum_{l=0\atop \scriptstyle{\rm even}}^{k-2}
C^{3/2}_{k-l-2}\left(\frac{x}{\xi}\right)\ P_l\left(\frac
1\xi\right)\int_{0}^1 dy\ y^{k-l-2}\ Q_k(y,t)\Biggr\}\, ,
\ee
with $H^{(k)}$ given by eq.~(\ref{kernel-podrobno}).\\
\underline{Amplitudes}\\
The amplitude with $C=-1$ exchange in the $t$-channel (odd signature) defined as

\be
A(\xi,t)=\int_{0}^1 dx\,H(x,\xi,t)\,
\left[\frac 1{x-\xi+i0} -\frac 1{x+\xi-i0}\right],
\ee

can be expressed in terms of the nonsinglet functions $Q_k(x,t)$
as follows:

\begin{eqnarray}
\nonumber
{\rm Re\ } A(\xi,t)&=&
-\int_0^{\frac{1-\sqrt{1-\xi^2}}{\xi}}
\frac{dx}{x}\sum_{k=0}^{\infty} x^k Q_k(x,t)
\Biggl[
\frac{1}{\sqrt{1-\frac{2 x}{\xi}+x^2}} -
\frac{1}{\sqrt{1+\frac{2 x}{\xi}+x^2}}
\Biggr]  \\
\nonumber
&+&\int^1_{\frac{1-\sqrt{1-\xi^2}}{\xi}}
\frac{dx}{x}\sum_{k=0}^{\infty} x^k Q_k(x,t)
\frac{1}{\sqrt{1+\frac{2 x}{\xi}+x^2}}
\, ,\\
{\rm Im\ } A(\xi,t)&=&
-\int_{\frac{1-\sqrt{1-\xi^2}}{\xi}}^1
\frac{dx}{x}\sum_{k=0}^{\infty} x^k Q_k(x,t)
\Biggl[
\frac{1}{\sqrt{\frac{2 x}{\xi}-x^2-1}}
\Biggr]\, .
\label{REIMnn}
\end{eqnarray}
From these expressions we can derive analytically the results for
the small $\xi$ behaviour of the amplitude for the case of $Q_k(x)\sim
x^{-\alpha_k}$. The result is

\be
\nonumber
{\rm Im} A^{(k)}(\xi,t)\sim -\left(\frac{\xi}{2}\right)^{k-\alpha_k}\,
\frac {\Gamma(1/2)\Gamma(\alpha_k-k+1/2)}{\Gamma(\alpha_k-k+1)}\, \\
{\rm Re} A^{(k)}(\xi,t)\sim {\rm Im} A^{(k)}(\xi,t)\ {\rm cotan}\left(\frac{\pi (\alpha_k-k-1)}{2}\right)\,.
\ee
The first equation again coincides with the small $\xi$ behaviour
of $-\pi H^{(k)}(\xi,\xi)$
as it should be. The second equation reproduces the well-known result of general
dispersion relations for the odd signature scattering amplitudes
at high energies.

\section*{\normalsize\bf Appendix B}
Here we sketch  the derivation of the
eqs.~(\ref{integralrelation},\ref{kernel-podrobno},\ref{kernel}). The basic relation we
use is the following\footnote{We define the discontinuity as ${{\rm disc}_{z=x}}f(z)=\frac{1}{\pi}\left[
f(x-i 0)-f(x+i 0)\right]$.}:

\be
\label{basic}
{{\rm disc}_{z=x}}\frac 1 y \left(1+y\frac{\partial}{\partial y}\right) \int_{-1}^1 ds\ z_s^{-N}=
\theta\left(1-\frac{x^2}{\xi^2}\right)\ \left(1-\frac{x^2}{\xi^2}\right)
\xi^{-N} y^{N-1} C_{N-1}^{3/2}\left(\frac x \xi\right)\,.
\ee
Here
\be
z_s=2\frac{z-s\xi}{(1-s^2)y}\, ,
\ee
with $0<y<1$. To prove eq.~(\ref{basic}) we note that:

\be
{{\rm disc}_{z=x}}z_s^{-N}=(-1)^{N-1}\frac{1}{\Gamma(N)}\ \delta^{(N-1)}(x_s)\, ,
\ee
with

\be
x_s=2\frac{x-s\xi}{(1-s^2)y}\, .
\ee
This simple relation gives us

\be
{{\rm disc}_{z=x}}\int_{-1}^1ds\ z_s^{-N}=(-1)^{N-1}\theta\left(1-\frac{x^2}{\xi^2}\right)
\frac{y^N}{2^N\xi^N\Gamma(N)}\ \left(\frac{\partial}{\partial s}\right)^{N-1}(1-s^2)\Biggr|_{s=x/\xi}\, .
\ee
The step function in the above equation indicates that the zero of the $\delta$-function should be
inside the interval $-1<s<1$. Further with help of the identity:
\be
\left(1+y\frac{\partial}{\partial y}\right)  z_s^{-N}=(N+1)z_s^{-N}\, ,
\ee
and the Rodrigues formula for the Gegenbauer polynomials:

\be
(1-x^2)\ C_{N-1}^{3/2}\left( x\right)=(-1)^{N-1} \frac{N+1}{2^N\Gamma(N)}
\left(\frac{\partial}{\partial x}\right)^{N-1}(1-x^2)^N\, ,
\ee
we arrive to the eq.~(\ref{basic}).

Now we consider the following function:
\be
\label{Fk}
F^{(k)}(z,y)\,=\,
\frac 1{y}\,\biggl(1\,+\,y\frac{\partial}{\partial y}\biggr)\,
\int_{-1}^1 ds\,\xi^k z_s^{1-k}\,
\bigl[z_s^2\,-\,2z_s\,+\,\xi^2\bigr]^{-1/2}\, ,
\ee
and let us compute its discontinuity with help of eq.~(\ref{basic}) and generating function for the
Legendre polynomials:
\be
{{\rm disc}_{z=x}}F^{(k)}(z,y)=\,\biggl(1-\frac{x^2}{\xi^2}\biggr)
\theta \biggl(1-\frac{x^2}{\xi^2}\biggr)
\sum_lC_{k+l-1}^{\,3/2}\left(\frac{x}{\xi}\right)\,P_l\left(\frac{1}{\xi}\right)
\,y^{k+l-1}\, .
\ee
In this expression we immediately recognize the integral kernel for the formal expansion
$$
H^{(k)}(x)\,=\,\biggl(1-\frac{x^2}{\xi^2}\biggr)
\theta \biggl(1-\frac{x^2}{\xi^2}\biggr)\,
\sum_lC_{k+l-1}^{\,3/2}\left(\frac{x}{\xi}\right)\,P_l\left(\frac{1}{\xi}\right)\
B_{k+l-1,l},
$$
with
$$
B_{k+l-1,l}\,=\,\frac 23\,\int dy\, y^{k+l-1}Q_k(y).
$$
Now the trick is that we can compute the discontinuity of the function $F^{(k)}(z,y)$
 given by eq.~(\ref{Fk})
 in a different way.
Namely, we take contributions to the discontinuity from the cut $1-\sqrt{1-\xi^2}<z_s<1+\sqrt{1-\xi^2}$
and from the poles at $z_s=0$ for $k\geq 2$. The calculations are straightforward, the cut contribution
gives us the expression (\ref{kernel}), whereas the pole contribution provides us with the second line in
eq.~(\ref{integralrelation}). Further, analyzing the solutions of the algebraic equation $x_s^2-2x_s+\xi^2=0$
we arrive at eq.~(\ref{kernel-podrobno}).


\begin{thebibliography}{99}

\bibitem{ZEUS}
P.~R.~Saull  [ZEUS Collaboration], ``Prompt photon production and
observation of deeply virtual Compton  scattering,''
hep-ex/0003030.
\bibitem{H1}
C.~Adloff {\it et al.}  [H1 Collaboration],
Phys.\ Lett.\ B {\bf 517} (2001) 47
[arXiv:hep-ex/0107005].

\bibitem{Amarian}
A.~Airapetian {\it et al.}  [HERMES Collaboration],
Phys.\ Rev.\ Lett.\  {\bf 87} (2001) 182001
[arXiv:hep-ex/0106068].
\bibitem{JLAB}
S.~Stepanyan {\it et al.}  [CLAS Collaboration],
Phys.\ Rev.\ Lett.\  {\bf 87} (2001) 182002
[arXiv:hep-ex/0107043].

\bibitem{HERAmeson}
J.~Breitweg {\it et al.}  [ZEUS Collaboration],
Eur.\ Phys.\ J.\ C {\bf 2} (1998) 247
[arXiv:hep-ex/9712020].\\
C.~Adloff {\it et al.}  [H1 Collaboration],
Eur.\ Phys.\ J.\ C {\bf 13} (2000) 371
[arXiv:hep-ex/9902019].

\bibitem{HERMESmeson}
A.~Airapetian {\it et al.}  [HERMES Collaboration],
``Single-spin azimuthal asymmetry in exclusive electroproduction of pi+  mesons,''
arXiv:hep-ex/0112022.\\
A.~Airapetian {\it et al.}  [HERMES Collaboration],
Eur.\ Phys.\ J.\ C {\bf 17} (2000) 389
[arXiv:hep-ex/0004023].

\bibitem{Jirev}
X.~Ji, J.~Phys.~G {\bf 24}, 1181 (1998).


\bibitem{Radrev}
A.~V.~Radyushkin,
hep-ph/0101225.
\bibitem{GPV}
K.~Goeke, M.~V.~Polyakov and M.~Vanderhaeghen,
Prog.\ Part.\ Nucl.\ Phys.\  {\bf 47} (2001) 401
[arXiv:hep-ph/0106012].

\bibitem{Belitsky:2001ns}
A.~V.~Belitsky, D.~Muller and A.~Kirchner,
arXiv:hep-ph/0112108.

\bibitem{DM}
D. M\"uller, D. Robaschik, B. Geyer, F.M. Dittes, and J. Horejsi,
Fortschr.~Phys. {\bf 42}, 101 (1994).

\bibitem{Rad96}
A.~V.~Radyushkin,
Phys.\ Lett.\ B {\bf 385} (1996) 333
[arXiv:hep-ph/9605431].
\bibitem{Rad}
A.V. Radyushkin, Phys.~Rev.~D {\bf 59}, 014030 (1999);
Phys.~Lett.~B {\bf 449}, 81 (1999).


\bibitem{PW99}
M.V. Polyakov and C. Weiss, Phys.~Rev.~D {\bf 60}, 114017 (1999).
\bibitem{Ji}
X. Ji, Phys.~Rev.~Lett. {\bf 78}, 610 (1997);
Phys.~Rev.~D {\bf 55}, 7114 (1997).

\bibitem{Mus}
I.~V.~Musatov and A.~V.~Radyushkin,
Phys.\ Rev.\ D {\bf 61} (2000) 074027
[arXiv:hep-ph/9905376].


\bibitem{BMKS}
A.~V.~Belitsky, D.~Muller, A.~Kirchner and A.~Schafer,
hep-ph/0011314.
\bibitem{OlegRadon}
O.V.~Teryaev, hep-ph/0102303.
\bibitem{ER}
A.V.~Efremov and A.V.~Radyushkin, Phys.\ Lett.\ {\bf B 94} (1980) 245.
\bibitem{BL}
G.P.~Lepage and S.J.~Brodsky, Phys.\ Lett.\ {\bf B 87} (1979) 359.

\bibitem{Pagels}
H.~Pagels, Phys. Rev. {\bf 144} (1966) 1250.

\bibitem{Donoghue:2001qc}
A.~A.~Pomeransky, R.~A.~Senkov and I.~B.~Khriplovich,
Phys.\ Usp.\  {\bf 43} (2000) 1055
[Usp.\ Fiz.\ Nauk {\bf 43} (2000) 1129].


J.~F.~Donoghue, B.~R.~Holstein, B.~Garbrecht and T.~Konstandin,
Phys.\ Lett.\ B {\bf 529} (2002) 132
[arXiv:hep-th/0112237].


\bibitem{Bur}
M.~Burkardt,
Phys.\ Rev.\ D {\bf 62} (2000) 071503
[arXiv:hep-ph/0005108].

M.~Burkardt,
arXiv:hep-ph/0207047.

\bibitem{Ral}
J.~P.~Ralston and B.~Pire,
arXiv:hep-ph/0110075.

\bibitem{Die}
M.~Diehl,
arXiv:hep-ph/0205208.

\bibitem{MVP98}
M.V.~Polyakov,
Nucl.~Phys. {\bf B555}, 231 (1999).


\bibitem{Pet98}
V.~Y.~Petrov, P.~V.~Pobylitsa, M.V.~Polyakov, I.~B\"ornig, K.~Goeke,
and C.~Weiss, Phys.~Rev.~D  {\bf 57}, 4325 (1998).
\bibitem{Belitsky:1998pc}
A.~V.~Belitsky, B.~Geyer, D.~Muller and A.~Schafer,
Phys.\ Lett.\ B {\bf 421} (1998) 312
[hep-ph/9710427].

\bibitem{Kiv99}
N.~Kivel and L.~Mankiewicz,
Nucl.\ Phys.\ B {\bf 557} (1999) 271
[hep-ph/9903531].

\bibitem{Kiv}
N.~Kivel, M.~V.~Polyakov and M.~Vanderhaeghen,
hep-ph/0012136.
\bibitem{Shu}
A.~G.~Shuvaev, K.~J.~Golec-Biernat, A.~D.~Martin and M.~G.~Ryskin,
Phys.\ Rev.\ D {\bf 60} (1999) 014015
[hep-ph/9902410].
\bibitem{Andrei}
A.~Shuvaev,
Phys.\ Rev.\ D {\bf 60} (1999) 116005
[hep-ph/9902318].
\bibitem{Gribov}
V.~N.~Gribov,
Nucl.\ Phys.\  {\bf 22} (1961) 249.

\bibitem{Nor}
J.~D.~Noritzsch,
Phys.\ Rev.\ D {\bf 62} (2000) 054015
[arXiv:hep-ph/0004012].

\bibitem{FreMcD}
A.~Freund and M.~McDermott,
Eur.\ Phys.\ J.\ C {\bf 23}, 651 (2002)
[arXiv:hep-ph/0111472].

A.~Freund and M.~F.~McDermott,
Phys.\ Rev.\ D {\bf 65}, 074008 (2002)
[arXiv:hep-ph/0106319].

A.~Freund and M.~F.~McDermott,
Phys.\ Rev.\ D {\bf 65}, 091901 (2002)
[arXiv:hep-ph/0106124].

\bibitem{evol}
M.~Gluck and E.~Reya,
Phys.\ Rev.\ D {\bf 14} (1976) 3034.

M.~Gluck, E.~Reya and A.~Vogt,
Z.\ Phys.\ C {\bf 48} (1990) 471.

D.~A.~Kosower,
Nucl.\ Phys.\ B {\bf 506} (1997) 439
[arXiv:hep-ph/9706213].

S.~Weinzierl,
arXiv:hep-ph/0203112.

\bibitem{Terpos}
B.~Pire, J.~Soffer and O.~Teryaev,
Eur.\ Phys.\ J.\ C {\bf 8} (1999) 103
[arXiv:hep-ph/9804284].
\bibitem{Radpos}
A.~V.~Radyushkin,
Phys.\ Rev.\ D {\bf 59} (1999) 014030
[arXiv:hep-ph/9805342].
\bibitem{Diehlpos}
M.~Diehl, T.~Feldmann, R.~Jakob and P.~Kroll,
Nucl.\ Phys.\ B {\bf 596} (2001) 33
[Erratum-ibid.\ B {\bf 605} (2001) 647]
[arXiv:hep-ph/0009255].


\bibitem{Pasha}
P.~V.~Pobylitsa,
arXiv:hep-ph/0204337.
P.~V.~Pobylitsa,
arXiv:hep-ph/0201030.

\bibitem{DiePirTer}
M.~Diehl, T.~Gousset, B.~Pire and O.~Teryaev,
Phys.\ Rev.\ Lett.\  {\bf 81} (1998) 1782
[arXiv:hep-ph/9805380].


\bibitem{BelMul}
A.~V.~Belitsky and D.~Muller,
Phys.\ Lett.\ B {\bf 417} (1998) 129
[arXiv:hep-ph/9709379].

A.~V.~Belitsky, A.~Freund and D.~Muller,
Nucl.\ Phys.\ B {\bf 574} (2000) 347
[arXiv:hep-ph/9912379].



\end{thebibliography}
\end{document}